\documentclass{sig-alternate}

\usepackage{listings}

\usepackage[hidelinks]{hyperref}
\usepackage{url}
\newcommand{\code}[1]{\texttt{#1}}
\newcommand{\floor}[1]{\left\lfloor #1 \right\rfloor}

\usepackage{amsmath}
\newtheorem{definition}{Definition}
\DeclareRobustCommand{\ttfamily}{\fontencoding{T1}\fontfamily{lmtt}\selectfont}

\begin{document}
%
\conferenceinfo{HPTCDL}{November 17, 2014, New Orleans, Louisiana, USA}
\CopyrightYear{2014}

\title{Parallel Prefix Polymorphism Permits \\
Parallelization, Presentation \& Proof}

\numberofauthors{2}

\author{
\alignauthor
Jiahao Chen\\
       \affaddr{Massachusetts Institute of Technology}\\
       \affaddr{Computer Science and Artificial Intelligence Laboratory}\\
       \affaddr{77 Massachusetts Avenue}\\
       \affaddr{Cambridge, Massachusetts 02139, USA}\\
       \email{jiahao@mit.edu}\\
\alignauthor
Alan Edelman\\
       \affaddr{Massachusetts Institute of Technology}\\
       \affaddr{Department of Mathematics, and Computer Science and Artificial Intelligence Laboratory}\\
       \affaddr{77 Massachusetts Avenue}\\
       \affaddr{Cambridge, Massachusetts 02139, USA}\\
       \email{edelman@mit.edu}
}

\date{15 October 2014}

\maketitle

\begin{abstract}
Polymorphism in programming languages enables code reuse. Here, we show that
polymorphism has broad applicability far beyond computations for technical
computing: \textbf{parallelism} in distributed computing, \textbf{presentation}
of visualizations of runtime data flow, and \textbf{proofs} for formal
verification of correctness. The ability to reuse a single codebase for all
these purposes provides new ways to understand and verify parallel programs.
\end{abstract}




\section{Introduction}

Abstractions are the lifeblood of computer science~\cite{Abelson1996}.
Separating higher level concepts from lower level implementation details allows
code to be more composable and reusable, and code using appropriate
abstractions is simpler to understand and maintain. However, the desire for
high level abstractions often seems inimical to writing high-performance code;
oftentimes performance can only result from code that takes advantage of
low-level implementation details and hence break abstractions. Thus areas like
technical computing and parallel programming, where performance is a key
consideration, exhibit a general dearth of abstraction and suffer the
consequent complexities in large, inscrutable and nearly unmaintainable
codebases. Such challenges are particularly acute for parallel programs, which
are written primarily for performance reasons. However, parallel programs are
notoriously prone to subtle, nondeterministic bugs arising from race
conditions, and they are difficult to verify for correctness. 

Programs for technical computing tend to sacrifice abstraction for performance,
but pay the cost in expressiveness. In contrast, the mathematical structure of
technical computations is inherently amenable to abstraction, particularly
polymorphism. Just think of the myriad ways different mathematical objects can
be multiplied together: one can multiply two numbers together, or two matrices
together, or a number and a matrix. The notion of multiplication itself can
also be extended to dot products, cross products, tensor products, wedge
products, outer products, and smash products, to just name a few. All these
operators share a common metaphor and it is natural to implement these products
with polymorphic code that can express the mathematical commonality.

In this paper, we argue that high level languages like Julia provide a suitable
framework for expressing high level abstractions that allow for extremely
powerful code reuse, while affording a reasonable level of performance.
In particular, we explore how Julia's generic function system and type system
are suitable for expressing polymorphism at the operator level, and show
how polymorphism can be used not just to encompass different kinds of
computations, but can be applied to parallelize code, enable visualizations of
code implementations, and even verify correctness of code.

\subsection{Multimethods in Julia}

In this paper, we use the Julia language\footnote{Julia is MIT-licensed open
source software and can be downloaded freely from \url{julialang.org}. We use
v0.3.1 in this paper.}, a very high level dynamic language designed
specifically for technical computing~\cite{Bezanson2012}. Julia offers language
constructs that support different mechanisms of polymorphism, which programming
language theorists call universal polymorphism and \textit{ad hoc}
polymorphism, or overloading~\cite{Strachey2000}. In this paper, we focus only
on Julia's overloading mechanism provided by multimethods.

\textit{Ad hoc} polymorphism, or overloading, is a language construct that
naturally expresses the polymorphism inherent in the mathematical structure of
technical computing~\cite{Bezanson2014}. Consider the \code{*} operator which
represents multiplication: the product can be taken between two integers, two
floating-point numbers, a scalar and a vector, or a matrix and a matrix, just
to list a few typical possibilities. All these different semantics can be
represented by with the same syntax, namely an expression of the form
\code{a*b}. All programming languages resolve the ambiguity by considering the
types of the arguments \code{a} and \code{b}, which is formally equivalent to
specifying the domains of the operands~\cite{Scott1976}.

In practice, programming languages vary greatly in how they allow users to
reason about types. Some languages, like Julia, offer the ability to define
multimethods, where a single generic function like \code{*} can be defined with
more than one method, each with a different type signature: in Julia
notation, \code{*(a::Number, b::Number)} defines a method for scalar
multiplication, whereas \code{*(a::Matrix, b::Vector)} defines a method for
matrix-vector products, and so on. Closely related is the support for multiple
dispatch, where the method chosen to match an expression like \code{a*b} can
depend on the type of more than one argument. In this respect, multimethods
differs greatly from more conventional languages that provide class-based
objects; dispatch only occurs on the first argument, which is the type of the
class.

In this paper, we demonstrate how multimethod polymorphism is far more general
than just dispatching on computational kernels. We study one specific
algorithm, namely scan, and show how the same exact code written in Julia for
serial computation can be composed with appropriately overloaded operators to
generate parallel code, visualizations, and also proofs of correctness.

\subsection{The scan algorithm}
\label{sec:prefix}

The basic problem of interest is to compute from some initial data \code{y} the
partial sums \code{z} such that:

\begin{verbatim}
z[1] = y[1]
z[2] = y[1] + y[2]
z[3] = y[1] + y[2] + y[3]
...
\end{verbatim}
One way to compute this sum efficiently is to note the prefix property, i.e.\
that the $k$th partial sum depends only on the $(k-1)$th partial sum and the
$k$th element: 

\begin{verbatim}
z[1] = y[1]
z[2] = z[1] + y[2]
z[3] = z[2] + y[3]
...
\end{verbatim}
which leads to the simple algorithm:

\begin{verbatim}
function prefix_serial!(y, +)
    for i=2:length(y)
        y[i] = y[i-1] + y[i]
    end
    y
end
\end{verbatim}

The cumulative sum problem generalizes to any associative operator; in this
Julia function, the argument \code{+} specifies the operator of interest,
allowing the same code to be reused for other operators like multiplication
(\code{*}), maximization (\code{max})~\cite{Shah2013}, or even string
concatenation\footnote{As written, the \code{prefix\allowbreak\_serial!}
function assumes, but does not check, that the function passed to it is
associative. If necessary, checks of the form

@assert (y[1]+y[2])+y[3] == y[1]+(y[2]+y[3])
can be included, but for simplicity of presentation, we omit such checks from
the code presented in this paper. We also neglect concerns relating to
\textit{approximate} associativity, such as roundoff errors in floating-point
addition or multiplication~\cite{Mathias1995}.}. The \code{!} suffix is a Julia
convention denoting that the function mutates at least one of its arguments; in
this case, the cumulative sums are computed in-place on \code{y}.

The general problem is called the prefix sum~\cite{Blelloch1989,Blelloch1993} or 
scan~\cite{Iverson1962,Iverson1979}. Nominally, it appears that the data has to
be scanned in one sweep from first to last and is a naturally serial process.
However, the insight behind parallel prefix
algorithms~\cite{Blelloch1989,Brent1982,Kogge1973,Kruskal1985,Ladner1980,Sklansky1960}
is that associativity allows the operations to regrouped in different ways
which can expose potential for concurrent execution, which can be interpreted
as generic divide-and-conquer strategies for recursive computation~\cite{Smith1987}.

In its general form, scan algorithms can be computed as a higher-order function
which takes as input some associative operator. Table~\ref{tab:prefixapps}
shows a representative list of applications of parallel prefix, showing the
diversity of applications and associative operators associated with those
applications~\cite{Blelloch1990,Blelloch1993}. Scan is therefore a prime
example of an algorithm that can exploit polymorphism for genericity.

\begin{table}
	\begin{tabular}{l l l}
		\hline
		Application              & Operator                  \\ \hline
		                         & Addition\\
		Poisson random variates \cite{Lu1996} & sequence lengths \\
		Minimal coverings \cite{Moitra1991}   & joining 2D regions \\
		Stream reduction \cite{Horn2005}      & counting records \\ \hline
	                                 & Maximization\\
		Line of sight \cite{Blelloch1990}          & height\\
		String alignment \cite{Hillis1986,Chi1992} & substring length\\ \hline
					 & Multiplication\\
		Binary addition \cite{Sklansky1960} 	      & Boolean matrices\\
		Polynomial interpolation \cite{Egecioglu1990} & scalars\\
		Sorting	\cite{Hillis1986,Blelloch1989} 		 & permutations\\
		Tridiagonal equations  \cite{Mathias1995}     & matrices\\ \hline
					 & Function composition & \\
		Finite state automata \cite{Ladner1980,Hillis1986}  & transition functions\\ \hline
	\end{tabular}
	\caption{Representative applications of the scan algorithm, employing
	four basic types of operations: addition, maximization, multiplication,
	and function composition.}
	\label{tab:prefixapps}
\end{table}

\subsection{The Brent--Kung form of parallel prefix}

In this paper, we focus on the Brent--Kung form~\cite{Brent1982} of parallel
prefix, where the computation is organized into two trees. For simplicity, we
present first the special case of parallel prefix for $n=8$ data points.

\begin{verbatim}
function prefix8!(y, +)
    length(y)==8 || error("length 8 only")
    for i in [2,4,6,8] y[i] = y[i-1] + y[i] end
    for i in [  4,  8] y[i] = y[i-2] + y[i] end
    for i in [      8] y[i] = y[i-4] + y[i] end
    for i in [    6  ] y[i] = y[i-2] + y[i] end
    for i in [ 3,5,7 ] y[i] = y[i-1] + y[i] end
    y
end
\end{verbatim}

\begin{figure}
  \centering

  \begin{verbatim}
  render(prefix_serial!(AccessArray(8),+))
  \end{verbatim}
  \includegraphics{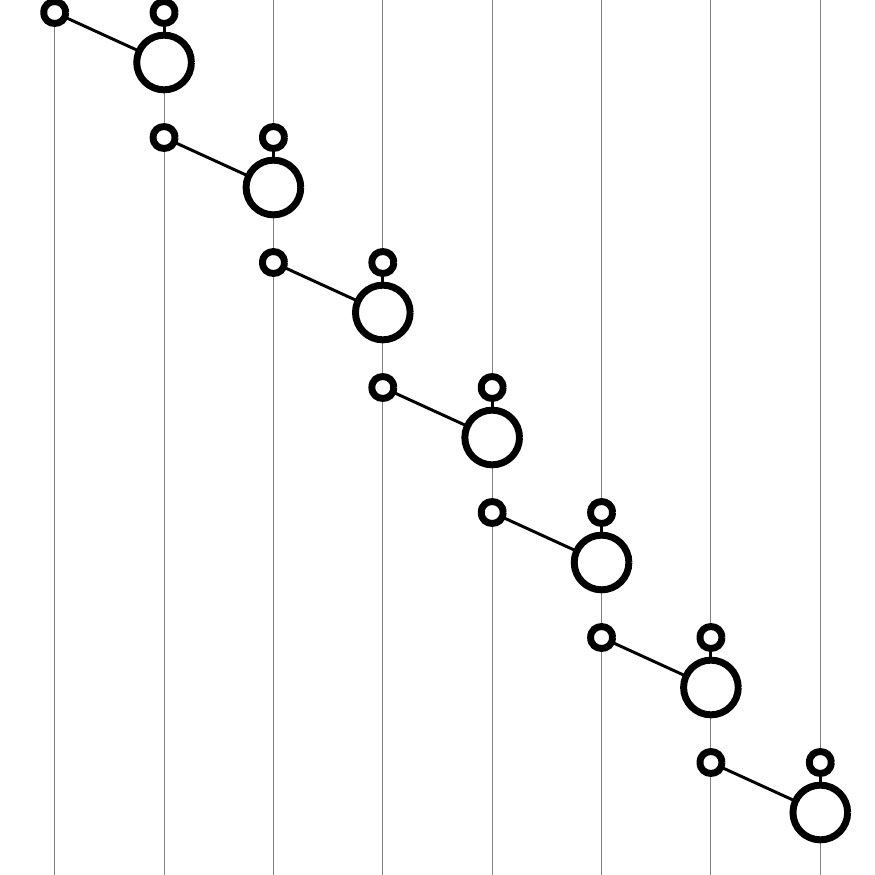}
  \vspace{12 pt}
  \begin{verbatim}
  render(prefix!(AccessArray(8),+))
  \end{verbatim}
  \includegraphics{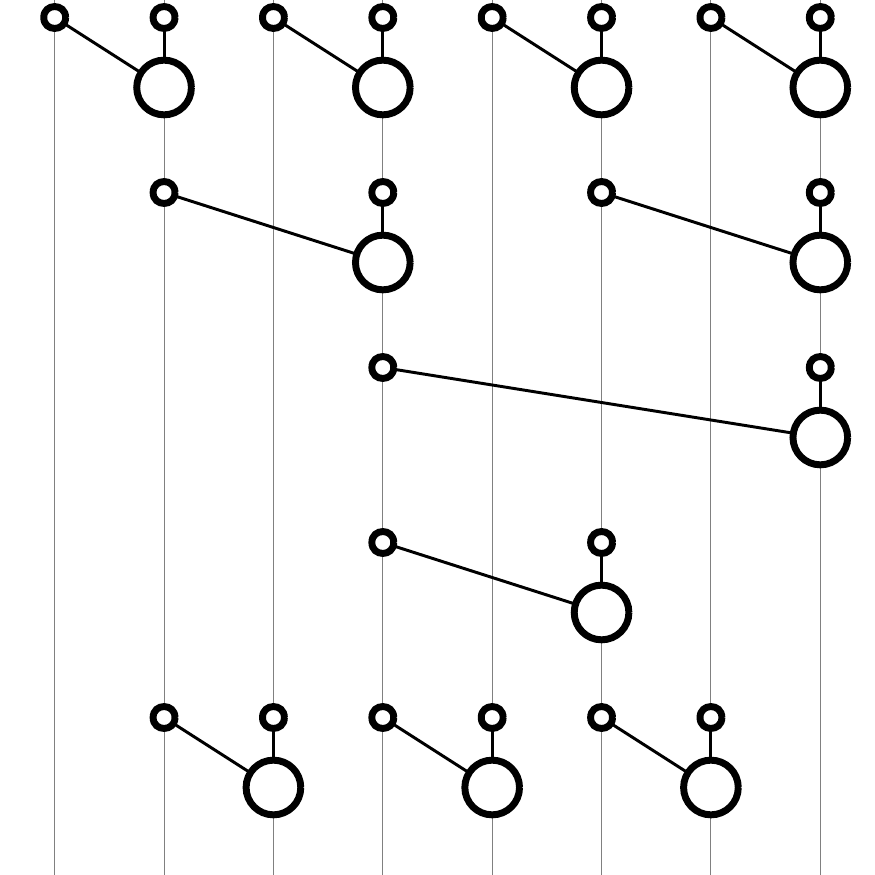}
  \caption{Above: operation order generated from the left-associative algorithm
	  \code{prefix\allowbreak\_serial!}.
	  Below: operation order generated from the tree algorithm \code{prefix8!}.
	  The figures were generated directly from the same kernels used for
	  computation in Section~\ref{sec:parallel-prefix}. The code listing
	  for the \code{render} function is given in Section~\ref{sec:render}.
	  This figure was rendered in Compose, a Julia package for declarative
          vector graphics~\cite{Compose.jl}.}
   \label{fig:gates}
\end{figure}

Figure~\ref{fig:gates} illustrates the difference between the number and order
of operations in \code{prefix\allowbreak\_serial!} and \code{prefix8!}. Each vertical line
represents a processor \code{i} operating on the data \code{y[i]}. Each
operation of the form \code{y[i] = y[j] + y[i]} is represented by a gate with
inputs on lines \code{i} and \code{j} and a single output on line \code{i}. The
main idea is that even though it takes more operations to organize the
computation in the double tree form of \code{prefix8!}, it is possible to
execute each stage of the computation tree concurrently, and parallel speedup
can be achieved if the depth of the resulting tree is shorter than the depth of
the tree for the serial algorithm. Nevertheless, at this point we have not
actually computed anything in parallel, merely organized the computation in a
way that would \textit{allow} for concurrent execution. Running the code as is
on an \code{Array} object would run the operations sequentially, from left to
right, then top to bottom of the computation tree. 

To conclude our exposition of the scan problem, we present the \code{prefix!}
function that solves the general case of $n$ data points. While the indices are
somewhat less clear than when explicitly written out in \code{prefix8!}, the
\code{prefix!} function nonetheless preserves the double tree structure.

\begin{verbatim}
function prefix!(y, +)
    l=length(y)
    k=iceil(log2(l))
    #The "reduce" tree
    for j=1:k, i=2^j:2^j:min(l, 2^k)
        y[i] = y[i-2^(j-1)] + y[i]
    end
    #The "broadcast" tree
    for j=(k-1):-1:1, i=3*2^(j-1):2^j:min(l, 2^k)
        y[i] = y[i-2^(j-1)] + y[i]
    end
    y
end
\end{verbatim}

Again, at this point we have only written serial code that introduces more
computations than the naive algorithm \code{prefix\allowbreak\_serial!}.
However, we will argue in Section~\ref{sec:parallel-prefix} that the exact same
code in \code{prefix!} can be reused for parallel execution which can achieve
speedup over \code{prefix\allowbreak\_serial!}.

\section{Operators for distributed computations}

In this section we show how the prefix algorithm we wrote above can be run in a
distributed setting without modification. The key is to make use of overloading
using the multimethod dispatch feature of Julia.

Julia provides native support for multiprocess distributed computing based on
one-sided message passing. The basic functionality is provided by the
\code{remotecall} function, which initiates a nonblocking remote function call
and returns an explicit future~\cite{Friedman1976} (a remote pointer of type
\code{RemoteRef}) whose value is retrieved by the \code{fetch} function, which
is a blocking operation. Julia also provides more convenient syntax for
\code{remotecall} with the \code{@spawn} and \code{@spawnat} macros, which
automatically rewrite Julia expressions into \code{remotecall} function calls.

We can use Julia's multiple dispatch feature to define associative operators
which act on remote data rather than local data. Julia's generic function
system allows new methods which act on remote data to be defined for functions
like \code{+} and \code{*}, which are simply functions for which the parser
supports infix notation. In effect, we can overload addition and multiplication
(or in general any binary associative function) transparently to work on remote
data.

For example, we can run the following code:

\begin{verbatim}
#Start a Julia process on every available core
#addprocs(n) adds n processors
#Sys.CPU_CORES is the total number of available
#CPU cores
#nprocs() returns the total number of Julia
#processes attached to the current master
#(including itself)
addprocs(max(0, Sys.CPU_CORES-nprocs()))

import Base.* #Extend existing generic function

#Define elementary operations on remote data
*(r1::RemoteRef,r2::RemoteRef)=
    @spawnat r2.where fetch(r1)*fetch(r2)
\end{verbatim}
This one method defines multiplication on remote data by \code{fetch}ing the
remote data from the process containing the data of \code{r1}, copying the data
of \code{fetch(r1)} to the memory space of the process with id \code{r2.where},
which already stores the data of \code{r2}. The process \code{r2.where} now
contains local copies of both operands. Assuming that the local data are of
type \code{T}, the Julia code then invokes another round of method dispatch
based on the method signature \code{*(::T, ::T)}. In this way, any data type
\code{T} that supports multiplication will now also support remote
multiplication, regardless of whether the data are scalar numbers, $N\times N$
matrices, or something else entirely.

The main point of this paper is that the very same function \code{prefix!}
which was executed in serial in previous sections will now run in parallel,
simply by passing to it an associative operator over remote data rather than
local data. Julia's multimethods and multiple dispatch semantics allow
operations on remote data to share the same syntax as their corresponding
operations on local data, thus removing any syntactic difference between remote
and local operations. The new method for \code{*} defines new behavior
specific to \code{RemoteRef}s, which are Julia's explicit futures. With this
new method defined in the current scope, running \code{prefix!(y, *)} will
automatically compute cumulative products on remote data if \code{y} is an
array of \code{RemoteRef}s. Julia will automatically dispatch on the
\code{*(r1::RemoteRef, r2::RemoteRef)} method within the inner loops of
\code{prefix!} by comparing the types of the data elements of \code{y} with
method signatures defined for \code{*}.

\subsection{Parallel prefix}
\label{sec:parallel-prefix}

We now run the \code{prefix!} function in parallel. The remote operations
\code{*(r1::RemoteRef, r2::RemoteRef)} contain blocking operations implied by
\code{fetch(r1)}, and Julia dynamically schedules all remote operations
simultaneously so long as they are not waiting on the result of a \code{fetch}
operation. The scheduling and dependency structure of \code{prefix!} thus
results in all operations in each stage of the tree being executed
simultaneously. Neglecting overhead from communication latency and bandwidth,
the total execution time of \code{prefix!} depends only on the depth of the
trees defined by the inner loops of \code{prefix!} and visualized in
Figure~\ref{fig:gates}.

From the indices of each loop in \code{prefix!} for $l$ data points, the first
tree has at least one operation at depth $k$ for $l \ge 2^k$, and therefore the
depth of the entire tree is $k = \floor{\log_2 l}$. Similarly, the second tree
has at least one operation at depth $k$ for $l \ge 3\cdot2^{k-1}$, and hence
has depth $k = 1 + \floor{log_2 \frac l 3}$. Adding these depths and assuming
that we distribute one datum per processor, we therefore obtain the theoretical
speedup ratio for $p$ processors running \code{prefix!} over
\code{prefix\allowbreak\_serial!} as:

\begin{equation}
    r (p) = \frac {p-1} {\floor{\log_2 p} + 1 + \floor{\log_2 \frac p 3}}.
    \label{eq:scaling-theory}
\end{equation}

\begin{figure}
  \centering
  \includegraphics[width=0.9\columnwidth]{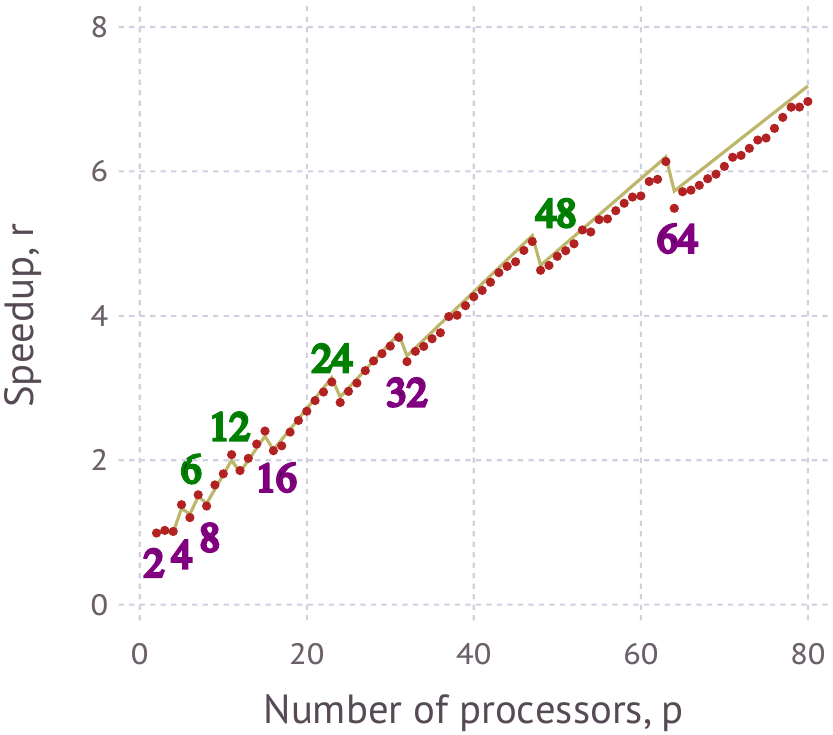}
  \caption{Weak scaling of the prefix sum kernels. Speedup ratios are the
  timings for \code{prefix!} over \code{prefix\allowbreak\_serial!}. Plotted as
  a solid line is the theoretical speedup ratio $r(p)$ of
  Equation~\ref{eq:scaling-theory}. This figure was rendered in Gadfly, a
  Julia package for native plotting and visualization~\cite{Gadfly.jl}.}
  \label{fig:scaling}
\end{figure}

Figure~\ref{fig:scaling} summarizes benchmark timings for a sample problem
where we generated $p$ square random matrices with Gaussian entries of size $n
= 4096$ and timed how long it took to multiply these matrices together on an
80-core Xeon E7-8850 machine with 1TB of shared memory. We specifically left
out the time needed to broadcast the data to the remote processes, so as to
focus only on the execution times of the kernels of interest. We also took care
to disable the garbage collector. Julia, like many high-level dynamic
languages, provides a garbage collector to aid in memory management. Julia
v0.3.1 uses a simple stop-the-world, non-moving, precise mark and sweep garbage
collector, where deallocation and finalization of garbage objects may not
happen immediately after objects become unused\footnote{The
code for Julia's garbage collector may be found at
\url{https://github.com/JuliaLang/julia/blob/275afc8b74b9c6ea5d34aefb8085525ff5dfc239/src/gc.c}}~\cite{McCarthy1960}.
Therefore, it becomes important to factor out the possible effects of
stop-the-world garbage collection. We explicitly disabled garbage collection
with \code{gc\_disable()} before running each kernel, then re-enabled garbage
collection with \code{gc\_enable()} after running each kernel. As an additional
precaution, we timed the kernels multiple times and took the minimum time for
each kernel so as to reduce fluctuations due to general nondeterministic
delays.

The empirical timings shown in Figure~\ref{fig:scaling} show excellent
agreement with the theoretical prediction of Equation~\ref{eq:scaling-theory},
with slight deterioration for $p>40$ cores reflecting the increased
communication overhead. The steps in the graph are as predicted by theory,
arising from the depth of the computation tree growing by one to accommodate
the extra data.

\subsection{Other variants of parallel prefix}

A wide variety of parallel prefix algorithms exist beyond the Brent--Kung form
~\cite{Blelloch1989,Egecioglu1992,Kogge1973,Kruskal1985,Ladner1980,Ofman1963,Sklansky1960,Sanders2006,Sengupta2007,Wang1996}.
All of these variants can also be expressed as generic higher-order functions
analogously to \code{prefix!}; however, not all of them can be written as
in-place computations~\cite{Merrill2009}. Nevertheless, the general principle
still holds that generic kernels can be written for an arbitrary parallel
prefix computation tree, and that these generic kernels can be composed
effortlessly in Julia to support exactly the same range of operations as the
Brent--Kung form in \code{prefix!}.

The genericity of \code{prefix!} is also very useful for implementing
so-called meta-strategies for parallel prefix such as
scan-then-fan~\cite{Merrill2009,Wilt2013}. The scan-then-fan variant of
parallel prefix is a divide-and-conquer algorithm, where parts of the scan are
chunked and computed serially on each process, after which each chunk is
offset by the correct result by the value of the last element in its preceding
chunk. The offsets can be propagated simply by running \code{prefix!} on the
distributed array of chunks, with the operator

\begin{verbatim}
+(a, b) = a[end] .+ b
\end{verbatim}
This variant of parallel prefix avoids the extra work incurred by generating a
computation tree that is wider than the total number of available processes,
and thus is more efficient to compute.

\section{Operator-level instrumentation}

Earlier, we showed in Figure~\ref{fig:gates} visualizations demonstrating the
double tree structure of the Brent--Kung parallel prefix algorithm and also the
cascading or rippling structure of the serial scan. These figures were
generated programmatically from the exact same kernels \code{prefix!} and
\code{prefix\_serial!} used to perform the computations.

Many visualizations of algorithms are bespoke; the representations are
completely decoupled from executable implementations. Alternatively, one may
envision generating visualizations of algorithms directly from code
implementations. Visualizations of algorithms can be generated by static
analysis: feed the compute kernel into another program as data to compute the
internal data flow. The static approach, however, is tantamount to
reimplementing the compiler to generate the correct execution trace, from which
the data flow can be inferred. Instead, one can employ dynamic analysis,
instrumenting the program much like a debugger. Conventional debuggers either
work on modified code with explicit instrumentation hooks embedded into the
original kernel, or run the program in a special virtual machine with
instrumentation hooks built into the low-level machine architecture. In these
dynamic analyses, the execution trace is reconstructed from the global machine
state, and again the data flow is inferred from the execution flow.

In this section, we describe a simple way to generate visualizations
programmatically by instrumenting the interface of specific data objects, namely
arrays. Instrumentation at this level retains the advantage of composing with
unmodified compute kernels, but does not require the sophisticated
infrastructure of an instrumented virtual machine, and reuses the static
analysis of the original compiler. Furthermore, the instrumentation occurs at
the level of individual variables, enabling highly selective traces
which are cheaper than conventional approaches which instrument the entire
program state. Additionally, the instrumentation measures the data flow
directly, rather than inferring it from global execution flow. The resulting
visualization provides an individual variable's point of view of what happens
over the course of a computation.

Our implementation in Julia takes advantage of genericity in the object model.
Unlike most traditional object-oriented paradigms, which focus on data
encapsulation~\cite{Cardelli1985}, the object model in Julia focuses on the
interface to objects provided by method calls~\cite{Mitchell1988}. Making the
behavior primary over the data contents lends more naturally to data
abstraction~\cite{Mitchell1988,Abadi1996}, and furthermore admits less conventional
object models involving multimethods and multiple dispatch~\cite{Castagna1997}.

\code{Array}s in Julia are containers of a given size (possibly with multiple
dimensions) and element type. The basic array interface for Julia provides size
and indexing semantics~\cite{Bezanson2014}. The basic interface is provided by
three functions:

\begin{description}
	\item[\code{length(A)}] returns the number of elements in the array
	      \code{A},
      	\item[\code{getindex(A, idx...)}] retrieves the element of the array
	      \code{A} with index \code{idx},
      	\item[\code{setindex!(A, val, idx...)}] puts the value \code{val} in
	      the array \code{A} at the index \code{idx}.
\end{description}

The Julia parser also provides syntax sugar for the latter two operations: code
like

\begin{verbatim}
A[i] = A[j] + A[k]
\end{verbatim}
is desugared into code of the form

\begin{verbatim}
x = getindex(A, j)
y = getindex(A, k)
z = x + y
setindex!(A, z, i)
\end{verbatim}

All the operations in the prefix sum kernels presented have array access
operations of this form: two \code{getindex} calls followed by one
\code{setindex!}. Based on this observation, we can write a very simple data
type that nominally provides exactly the same interface as an \code{Array}, but
rather than actually storing data elements, merely records the indices accessed
by indexing operations.

Here is the entire Julia code that implements \code{AccessArray}, an abstract
array type which instruments its indexing operations:

\begin{verbatim}
import Base: getindex, setindex!, length

type AccessArray
    length :: Int
    read :: Vector
    history :: Vector
    AccessArray(length)=new(length, Any[], Any[])
end

length(A::AccessArray)=A.length

function getindex(A::AccessArray, i)
    push!(A.read, i)
    nothing
end

function setindex!(A::AccessArray, x, i)
    push!(A.history, (A.read, Any[i]))
    A.read = Any[]
end

#Dummy associative operator
+(a::Void, b::Void) = nothing
\end{verbatim}
The \code{AccessArray} type contains three fields:

\begin{description}
	\item[\code{length}] the effective length of the array,
	\item[\code{read}] the history of indices accessed by \code{getindex}
		that have yet to be followed by a \code{setindex!} call, and
	\item[\code{history}] the history of indices accessed by
		(\code{getindex}, \code{set\-index!}) calls.
\end{description}
The \code{Any[]} construct defines an empty array which is explicitly typed to
allow elements of any type. \code{getindex} is defined to always return the value
\code{nothing}\footnote{\code{nothing} is a value of the special
singleton type \code{Void}, akin to Python's \code{none} or Haskell's
\code{Nothing}.}, while recording the index \code{i} into \code{A.read}.
\code{setindex!} records the index \code{i}, pairs it with the current value of
\code{A.read}, and stores the pair into \code{A.history}.

As implemented, the \code{AccessArray} type can only accurately trace code
where a \code{setindex!} call uses all the data from previous \code{getindex}
calls. Furthermore, it does not handle cases where execution flow depends on
the values of the array elements. Nevertheless, the \code{AccessArray} type is
sufficiently powerful to record transactions relevant for prefix sums, and can
be extended to more general execution flow patterns if necessary by wrapping
actual data elements. The Appendix further defines the \code{render} function
used to construct vector graphical objects from the instrumentation data stored
in an \code{AccessArray}, and Figure~\ref{fig:gates} shows graphical renderings
of the access patterns produced by \code{prefix\_serial!} and \code{prefix!}.

\section{Operators for formal verification}

In Section~\ref{sec:prefix} we introduced several different kernels to compute
scans. But how do we know that these kernels compute the prefix sum correctly?
Each of these kernels have exactly the same function signature \code{(y,
+)} representing the data \code{y} and associative binary operator \code{+}.
It turns out that the inputs \code{(y, +)} to the scan algorithm turn out to
have exactly the algebraic structure of a monoid, if the domain of array
elements \code{y[i]} contains an identity under the operation \code{+}. The
monoidal structure has been used in at least two ways to prove correctness.
First, \cite{Hinze2004} constructed a formal algebra that allows correctness of
circuits to be proved by derivation: all circuits which are equivalent to a
known correct circuit, up to certain algebraic transformations, will all be
correct. However, the algebraic proof of correctness is not constructive and
does not lend itself easily to programmatic verification. Second and more
recently, \cite{Chong2014} proved that the correctness of a kernel can be
demonstrated by proving correctness for the interval monoid
(Definition~\ref{def:intervalmonoid}), which formalizes the notion of indexing
the subarrays being accessed over the course of the prefix sum computation. The
latter method of proof is easy to verify programmatically.

In this section, we show how polymorphism allows the same Julia code written in
previous sections for practical computations to also be used in the formal
setting of verifying correctness. For convenience, we quote the definition of
the interval monoid:

\begin{definition}{\cite[Definition 4.3]{Chong2014}}
\label{def:intervalmonoid}
	
The \textit{interval monoid} $I$ has the elements

\begin{equation}
	\mathbb S_I = \left\{ (i_1, i_2) \in \mathrm{Int} \times \mathrm{Int} \;\vert\; i_1 \le i_2  \right\} \cup \{\mathbf 1_I, \top \}
\end{equation}
and a binary operator $\oplus_I$ defined by:

\begin{subequations}
\begin{align}
	\mathbf 1_I \oplus_I x = x \oplus_I \mathbf 1_I &= x \textrm{ for all } x \in \mathbb S_I \\
	\top \oplus_I x = x \oplus_I \top &= \top \textrm{ for all } x \in \mathbb S_I \\
	(i_1, i_2) \oplus_I (i_3, i_4) &= \begin{cases} (i_1, i_4) &\textrm{if } i_2 + 1 = i_3 \\
		\top &\textrm{otherwise.}
\end{cases}
\label{eq:intervalplus}
\end{align}
\end{subequations}

\end{definition}

The elements $(i, j) \in \mathbb S_I$ are abstractions of array indexing
operations which produce array slices; they are produced by Julia code like
\code{y[i:j]} where \code{i:j} is of type \code{UnitRange} and is a range of
unit stride representing the set $\{i, i+1, \dots, j\}$. The definition of
$\oplus_I$ in \eqref{eq:intervalplus} formalizes the notion of combining the
results from the subarrays \code{y[i:j]} and \code{y[j+1:k]} to get the result
for the subarray \code{y[i:k]}.  The identity element $\mathbf 1_I$ formalizes
an empty interval, while the annihilator $\top$ encodes noncontiguous ranges,
which correspond to partial sums which cannot be represented by slicing with a
\code{UnitRange}.

The key insight of \cite{Chong2014} is that correct computations of prefix sums
cannot generate noncontiguous elements $\top$, otherwise they would by
definition violate the prefixing property \code{prefix!(y[1:j+1], +) =
prefix!(y[1:j], +) + y[j+1]}.\\
From this insight, the authors of \cite{Chong2014} derive two correctness
results:

\begin{enumerate}
		
\item 	A function that computes the prefix sum in serial is correct for $n$
	data points if and only if that function computes the correct answer
	for the input \\
	$\left(\left((1, 1), (2, 2), \dots, (n, n)\right),
	\oplus_I\right)$\footnote{Our presentation differs from the original
	only in that Julia arrays are 1-based, in contrast to C/OpenCL arrays
	studied in the original~\cite{Chong2014}, which are 0-based.}
	~\cite[Theorem 4.5]{Chong2014}.
	Furthermore, the correct answer is $\left((1, 1), (1, 2), \dots, (1, n)
	\right)$, as the $k$th partial sum involves summing the subarray
	\code{y[1:k]}.
	
\item	A function that computes the prefix sum in parallel is correct if it is
	free of data races and its equivalent serialization is
	correct~\cite[Theorem 5.3]{Chong2014}.

\end{enumerate}

We can use these results directly to verify the correctness of the Julia code
we have written in earlier sections. By construction, the \code{fetch}es on
\code{RemoteRefs} insert implicit synchronization barriers and thus the
parallel code is free of data races. Thus only the serial correctness result
needs to be verified explicitly.

Julia allows us to encode the interval monoid directly from the definition, by
making use of the rich type system which is exposed to the user. The type
system is conventionally used in Julia for type inference and data abstraction;
here, we exploit the Curry--Howard correspondence to use the type system as a
computational resource that can be used to prove
correctness~\cite{Curry1958,Tait1965,Howard1980}. A convenient feature of
Julia's type system is the ability to use abstract data types as
singleton values: Julia types are values, and types can be used as singleton
values using the \code{Type\{T\}} construct. Thus, the domain $\mathbb S_I$ can
be written as a Julia type \code{S}, which is the \code{Union} (type union) of:

\begin{itemize}
	\item \code{UnitRange},
	\item \code{Type\{Id\}}, the identity singleton $\mathbf 1_I$, and
	\item \code{Type\{Tee\}}, the annihilator singleton $\top$.
\end{itemize}

With this mapping of the abstract interval monoid domain $\mathbb S_I$ onto
Julia types, Definition~\ref{def:intervalmonoid} translates directly into the
following code:

\begin{verbatim}
#S is the domain of the interval monoid, $\mathbb{S}_I$
abstract Tee #$\top$
abstract Id  #$\mathbf{1}_I$
typealias S Union(UnitRange, Type{Tee}, Type{Id})

#+ is the operator of the interval monoid, $\oplus_I$
+(I::UnitRange, J::UnitRange) = #$+_1$ 
    I.stop+1==J.start ? (I.start:J.stop) : Tee
+(::Type{Id}, ::Type{Id}) = Id  #$+_2$
+(I::S, ::Type{Id}) = I         #$+_3$
+(::Type{Id}, J::S) = J         #$+_4$
+(I::S, J::S) = Tee             #$+_5$
\end{verbatim}

\begin{figure}
  \centering
  \includegraphics[width=.9\columnwidth]{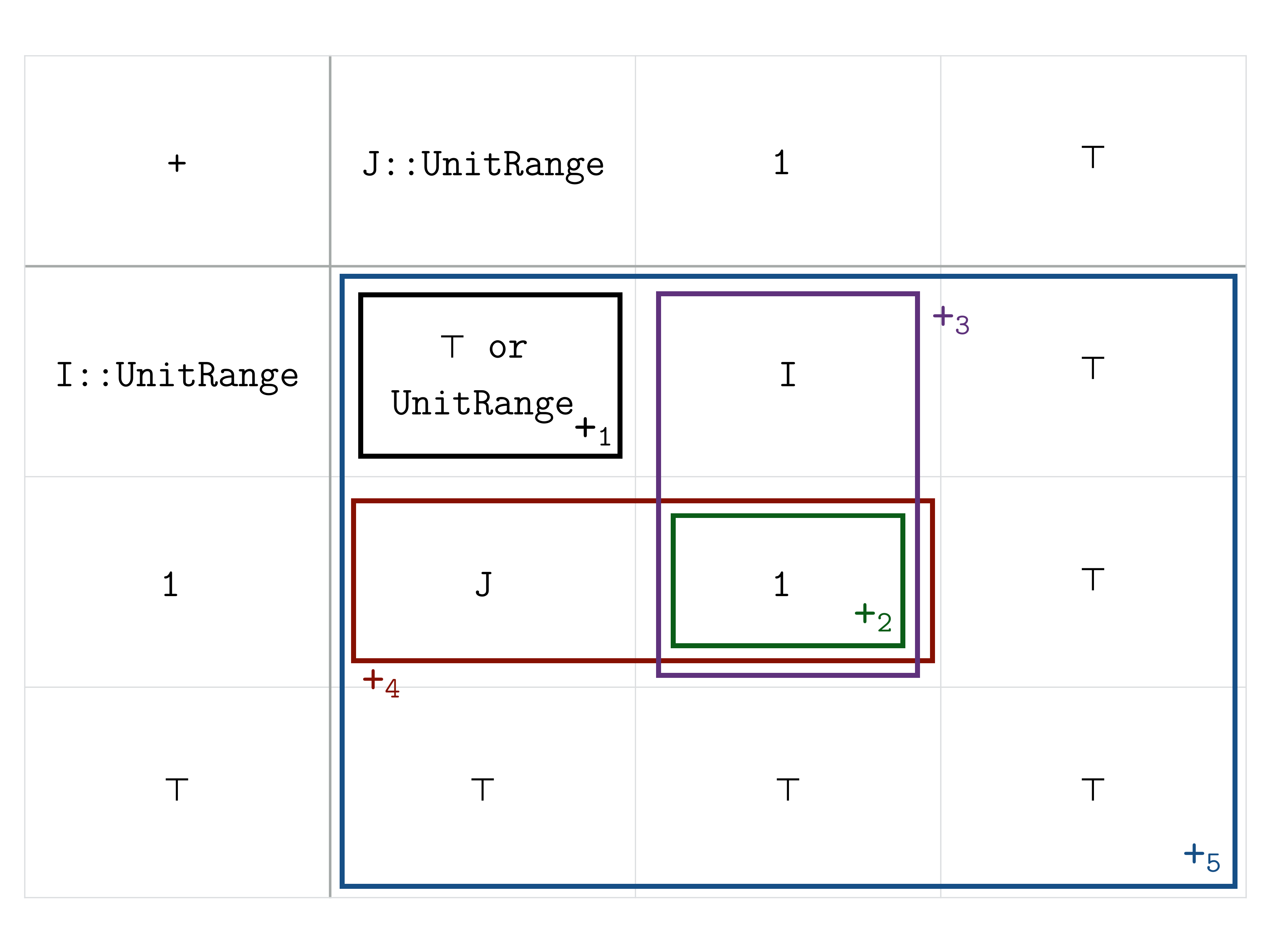}
  \caption{Operation table for the interval monoid $(\mathbb S_I, \oplus_I)$,
	  showing the overlapping domains of the various methods
	  $+_{i\in\{1,\dots,5\}}$. The dispatch rules in Julia choose the most
  	  specific method defined over the given cell~\cite{Bezanson2012}.}
  \label{fig:dispatch}
\end{figure}

Figure~\ref{fig:dispatch} summarizes the method dispatch table for the interval
monoid, which demonstrates the use of some interesting features of Julia's
method dispatcher~\cite{Bezanson2012}. First, the Julia method dispatcher
chooses the most specific method that matches the type signature of a given set
of arguments. Thus even though \code{+} may appear ambiguous for inputs of 
type \code{(::Unit\-Range, ::UnitRange)}, which matches both $+_1$ and $+_5$
methods, Julia resolves the ambiguity in favor of $+_1$ which has the more
specific type signature, since by definition \code{UnitRange} is a subtype of
\code{S}. Second, Julia uses symmetric multiple dispatch: the positions of the
arguments are not used to resolve ambiguities. Hence we need the special-case
method $+_2$ with type signature \code{(::Type\{Id\}, ::Type\{Id\})}, which
lies in the intersection of the type signatures of $+_3$ and $+_4$.  Bearing
these rules in mind, it is straightforward to verify that the definition of
\code{+} in the code block above is equivalent to that of $\oplus_I$ in
Definition~\ref{def:intervalmonoid}. Julia's method dispatch rules allow
\code{+} to be defined in a way that reveals the catch-all nature of $\top$:
method $+_5$, which returns \code{Tee}, is dispatched only when none of the
other methods matches the type signature of the given arguments.

Verifying some function \code{kernel} for the problem size \code{n} simply
reduces to writing the assertion:

\begin{verbatim}
#Test that kernel is correct for problem size n
@assert kernel([k:k for k=1:n],+)==[1:k for k=1:n]
\end{verbatim}

Attempting to verify an incorrect kernel results in at least one $\top$ being
produced during the computation, thus poisoning the program state and
precipitating type conversion errors of the form

\begin{verbatim}{jlcon}
`convert` has no method matching
convert(::Type{UnitRange}, ::Type{Tee})
\end{verbatim}
which arise from the inability of noncontiguous ranges to be expressed as
\code{UnitRange}s. 

The Curry--Howard correspondence allows us to verify correct programs as
programs with no type errors; programs with type errors must necessarily be
incorrect. Julia thus allows for the same kernels used for computation to be
verified directly without any rewriting or translation, simply by exploiting
the polymorphism arising from the generic nature of prefix sum kernels, and
composing such generic functions with appropriate choices of input data types
and associative operators over those types.

\section{Related work}

Julia does not provide parallel prefix in the base library; however, several
languages like APL~\cite{Iverson1962,Iverson1979}, Chapel~\cite{Deitz2006},
C**~\cite{Viswanathan1996} and ZPL~\cite{Chamberlain2000,Deitz2002} do provide
scan primitives. Other languages can use commonly-used library routines for
scans: MPI provides the \code{MPI\_scan} primitive~\cite{Snir1995,MPI}, and in
MPI-2, also the \code{MPI\_Exscan} primitive for exclusive scan~\cite{MPI2}.
Intel's Threading Building Blocks (TBB) library provides similar
functionality~\cite{Reinders2007}.  GPU-specific implementations also exist,
such in Haskell's Accelerate library~\cite{Chakravarty2011} and the Thrust C++
library~\cite{Bell2012}. Most of these implementations, however, either lack
genericity or express genericity through cumbersome language semantics. APL
does not provide generic parallel prefix, and parallelism is provided only by
nonstandard implementations. C** only supports user-definable parallel prefix
for commutative operations~\cite{Viswanathan1996}. ZPL allows only for a
limited form of overloading in terms of existing
operators~\cite{Deitz2002,Deitz2006}.  MPI allows parallel prefix on any
\code{MPI\_Datatype} and \code{MPI\_op}; user-defined operations and data types
can be used, but must be explicitly wrapped to do so, and furthermore requires
users to reason explicitly about low-level parallelism. The Haskell Accelerate
library provides genericity by generating code from user-specified expressions
into code that implements parallel prefix, but such code does not benefit from
Haskell's static type system since they are not statically analyzable. Thrust
and TBB, being written in C++, requires user-specified functions to be
specified as C++ functors, which are cumbersome to write as they must be
expressed using C++ expression templates. Chapel~\cite{Deitz2006} comes
closest to our work in providing user-definable scan operators in convenient
syntax; however, operators have to be explicitly defined as classes, which do
not support the full polymorphic expressiveness of multimethods.

Our implementation of parallel prefix as a higher-order function in Julia is
generic in that it makes use of duck typing: so long as the input operation is
associative and the collection of data is indexable, then the prefix sum
computation will simply work. The simplicity of this generic approach, however,
is by design rather naive and does not account for the complexities in real
world implementations, for example possible synchronicity issues produced
by higher levels of the broadcast and reduce trees that could result in bus
saturation. Nevertheless, we can handle resource constraints by using a more
sophisticated dynamic scheduler.

\section{Conclusions and outlook}

We have demonstrated how polymorphism using multimethods allows us to write
generic computation kernels for parallel prefix, which can then be composed
seamlessly with appropriate types and associative operators not only for
computation for a wide variety of problems, but also for various code
introspection tasks such as visualization and formal verification using the
interval monoid. Julia's language features lends to very natural and minimalist
code that takes full advantage of \textit{ad hoc} polymorphism to provide a
single set of computational kernels that can also be visualized directly and
verified without the need for retranslation and reimplementation in a more
specialized language.

\section{Acknowledgments}

The ideas in this paper were developed from examples in 18.337, the MIT course
on parallel computing, in the Fall 2013 semester. We gratefully acknowledge the
Julia community, especially Jeff Bezanson and Jake Bolewski, for insightful
discussions. Funding for this work was provided by the Intel Science and
Technology Center for Big Data, the MIT Deshpande Center Innovation Grant, the
DARPA XDATA Program, and the National Science Foundation Grant DMS-1312831.

\bibliographystyle{abbrv}
\bibliography{prefix}

\begin{thebibliography}{10}

\bibitem{Abadi1996}
M.~Abadi and L.~Cardelli.
\newblock {\em A theory of objects}.
\newblock Number~1 in Monographs in Computer Science. Springer-Verlag, New
  York, New York, USA, 1996.

\bibitem{Abelson1996}
H.~Abelson, G.~J. Sussman, and J.~Sussman.
\newblock {\em Structure and Interpretation of Computer Programs}.
\newblock MIT Press/McGraw-Hill, Cambridge, Massachusetts, 2 edition, 1996.

\bibitem{Bell2012}
N.~Bell and J.~Hoberock.
\newblock {T}hrust: A productivity-oriented library for {CUDA}.
\newblock In W.-M.~W. Hwu, A.~Schuh, N.~Mohammad, T.~Bradley, F.~Jargstorff,
  P.~Micikevicius, R.~Tonge, P.~Wang, and C.~Wooley, editors, {\em GPU
  Computing Gems Jade Edition}, Applications of GPU Computing, chapter~26,
  pages 359--371. Morgan-Kaufmann, Waltham, MA, 2012.

\bibitem{Bezanson2014}
J.~Bezanson, J.~Chen, S.~Karpinski, V.~Shah, and A.~Edelman.
\newblock Array operators using multiple dispatch: A design methodology for
  array implementations in dynamic languages.
\newblock In {\em Proceedings of ACM SIGPLAN International Workshop on
  Libraries, Languages, and Compilers for Array Programming}, ARRAY'14, pages
  56--61, New York, NY, USA, 2014. ACM.

\bibitem{Bezanson2012}
J.~Bezanson, S.~Karpinski, V.~B. Shah, and A.~Edelman.
\newblock {J}ulia: A fast dynamic language for technical computing.
\newblock {\em arXiv:1209.5145 [cs.PL]}.

\bibitem{Blelloch1989}
G.~E. Blelloch.
\newblock Scans as primitive parallel operations.
\newblock {\em IEEE Trans. Comput.}, 38(11):1526--1538, 1989.

\bibitem{Blelloch1990}
G.~E. Blelloch.
\newblock {\em Vector models for data-parallel computing}.
\newblock Artificial Intelligence. MIT Press, Cambridge, Massachusetts, 1990.

\bibitem{Blelloch1993}
G.~E. Blelloch.
\newblock Prefix sums and their applications.
\newblock In J.~H. Reif, editor, {\em Synthesis of Parallel Algorithms},
  chapter~1, pages 35--60. Morgan Kaufmann, San Mateo, California, 1993.

\bibitem{Brent1982}
R.~P. Brent.
\newblock A regular layout for parallel adders.
\newblock {\em IEEE Trans. Comput.}, C-31(3):260--264, Mar. 1982.

\bibitem{Cardelli1985}
L.~Cardelli and P.~Wegner.
\newblock On understanding types, data abstraction, and polymorphism.
\newblock {\em ACM Comput. Surveys}, 17(4):471--523, 1985.

\bibitem{Castagna1997}
G.~Castagna.
\newblock {\em Object-oriented programming: a unified foundation}.
\newblock Progress in Theoretical Computer Science. Birkh{\"a}user, Boston,
  1997.

\bibitem{Chakravarty2011}
M.~M.~T. Chakravarty, G.~Keller, S.~Lee, T.~L. McDonell, and V.~Grover.
\newblock Accelerating {H}askell array codes with multicore {GPU}s.
\newblock In {\em Proceedings of the Sixth Workshop on Declarative Aspects of
  Multicore Programming - DAMP '11}, pages 3--14, New York, New York, USA,
  2011. ACM Press.

\bibitem{Chamberlain2000}
B.~L. Chamberlain, S.-E. Choi, E.~C. Lewis, C.~Lin, L.~Snyder, and W.~D.
  Weathersby.
\newblock {ZPL}: a machine independent programming language for parallel
  computers.
\newblock {\em IEEE Trans. Software Eng.}, 26(3):197--211, Mar. 2000.

\bibitem{Chi1992}
L.~Chi and K.~Hui.
\newblock Color set size problem with applications to string matching.
\newblock In A.~Apostolico, M.~Crochemore, Z.~Galil, and U.~Manber, editors,
  {\em Combinatorial Pattern Matching}, Lecture Notes in Computer Science,
  chapter~19, pages 230--243. Springer, Berlin, Heidelberg, 1992.

\bibitem{Chong2014}
N.~Chong, A.~F. Donaldson, and J.~Ketema.
\newblock A sound and complete abstraction for reasoning about parallel prefix
  sums.
\newblock In {\em Proceedings of the 41st ACM SIGPLAN-SIGACT Symposium on
  Principles of Programming Languages - POPL '14}, pages 397--409, New York,
  New York, USA, 2014. ACM Press.

\bibitem{Curry1958}
H.~B. Curry.
\newblock {\em Combinatory Logic, Volume I}, volume~22 of {\em Studies in logic
  and the foundations of mathematics}.
\newblock North-Holland Publishing, Amsterdam, 2 edition, 1958.

\bibitem{Deitz2006}
S.~J. Deitz, D.~Callahan, B.~L. Chamberlain, and L.~Snyder.
\newblock {Global-view abstractions for user-defined reductions and scans}.
\newblock In {\em Proceedings of the eleventh ACM SIGPLAN symposium on
  Principles and practice of parallel programming - PPoPP '06}, pages 40--47,
  New York, New York, USA, 2006. ACM Press.

\bibitem{Deitz2002}
S.~J. Deitz, B.~L. Chamberlain, and L.~Snyder.
\newblock {High-level language support for user-defined reductions}.
\newblock {\em The Journal of Supercomputing}, 23(1):23--37, 2002.

\bibitem{Egecioglu1990}
{\"O}.~E{\u g}ecio{\u g}lu, E.~Gallopoulos, and {\c C}.~K. Ko{\c c}.
\newblock A parallel method for fast and practical high-order {N}ewton
  interpolation.
\newblock {\em BIT Numer. Math.}, 30(2):268--288, June 1990.

\bibitem{Egecioglu1992}
{\"O}.~E{\u g}ecio{\u g}lu and {\c C}.~K. Ko{\c c}.
\newblock Parallel prefix computation with few processors.
\newblock {\em Comput. Math. App.}, 24(4):77--84, 1992.

\bibitem{Friedman1976}
D.~Friedman and D.~Wise.
\newblock The impact of applicative programming on multiprocessing.
\newblock In {\em Proceedings of the 1976 International Conference on Parallel
  Processing}, pages 263--272, Long Beach, CA, 1976. IEEE.

\bibitem{MPI}
W.~Gropp, E.~Lusk, and A.~Skjellum.
\newblock {\em Using {MPI}: Portable Parallel Programming with the
  Message-Passing Interface}.
\newblock MIT Press, Cambridge, Massachusetts, 2 edition, 1999.

\bibitem{MPI2}
W.~Gropp, E.~Lusk, and R.~Thakur.
\newblock {\em Using {MPI}-2: Advanced Features of the Message-Passing
  Interface}.
\newblock MIT Press, Cambridge, Massachusetts, 1999.

\bibitem{Hillis1986}
W.~D. Hillis and G.~L. {Steele, Jr.}
\newblock Data parallel algorithms.
\newblock {\em Commun. ACM}, 29(12):1170--1183, Dec. 1986.

\bibitem{Hinze2004}
R.~Hinze.
\newblock An algebra of scans.
\newblock In D.~Kozen, editor, {\em Mathematics of Program Construction},
  Lecture Notes in Computer Science, chapter~11, pages 186--210. Springer,
  Berlin, Heidelberg, 2004.

\bibitem{Horn2005}
D.~Horn.
\newblock Stream reduction operations for gpgpu applications.
\newblock In M.~Pharr, editor, {\em GPU Gems 2}, chapter~36, pages 573--589.
  Addison-Wesley, 2005.

\bibitem{Howard1980}
W.~A. Howard.
\newblock The formulas-as-types notion of construction.
\newblock In J.~P. Seldin and J.~R. Hindley, editors, {\em To {H}. {B}.
  {C}urry: Essays on Combinatory Logic, Lambda Calculus, and Formalism}, pages
  479--490. Academic Press, New York, New York, USA, 1980.

\bibitem{Iverson1962}
K.~E. Iverson.
\newblock {\em A programming language}.
\newblock John Wiley \& Sons, New York, NY, USA, 1962.

\bibitem{Iverson1979}
K.~E. Iverson.
\newblock Operators.
\newblock {\em ACM Trans. Program. Lang. Sys.}, 1(2):161--176, Oct. 1979.

\bibitem{Gadfly.jl}
D.~C. Jones, D.~Chudzicki, A.~Sengupta, et~al.
\newblock Gadfly.jl v0.3.9 (gadflyjl.org) - native plotting and visualization
  for {J}ulia.

\bibitem{Compose.jl}
D.~C. Jones, D.~Darakananda, K.~Fischer, et~al.
\newblock Compose.jl v0.3.9 (composejl.org) - declarative vector graphics for
  {J}ulia.

\bibitem{Kogge1973}
P.~M. Kogge and H.~S. Stone.
\newblock A parallel algorithm for the efficient solution of a general class of
  recurrence equations.
\newblock {\em IEEE Trans. Comput.}, C-22(8):786--793, Aug. 1973.

\bibitem{Kruskal1985}
C.~P. Kruskal, L.~Rudolph, and M.~Snir.
\newblock The power of parallel prefix.
\newblock {\em IEEE Trans. Comput.}, C-34(10):965--968, 1985.

\bibitem{Ladner1980}
R.~E. Ladner and M.~J. Fischer.
\newblock Parallel prefix computation.
\newblock {\em J. ACM}, 27(4):831--838, 1980.

\bibitem{Lu1996}
T.-C. Lu, Y.-S. Hou, and R.-J. Chen.
\newblock A parallel {P}oisson generator using parallel prefix.
\newblock {\em Comput. Math. App.}, 31(3):33--42, Feb. 1996.

\bibitem{Mathias1995}
R.~Mathias.
\newblock The instability of parallel prefix matrix multiplication.
\newblock {\em SIAM J. Sci. Comput.}, 16(4):956--973, July 1995.

\bibitem{McCarthy1960}
J.~McCarthy.
\newblock Recursive functions of symbolic expressions and their computation by
  machine, {P}art {I}.
\newblock {\em Commun. ACM}, 3(4):184--195, Apr. 1960.

\bibitem{Merrill2009}
D.~Merrill and A.~Grimshaw.
\newblock Parallel scan for stream architectures.
\newblock Technical Report December 2009, Department of Computer Science,
  University of Virginia, Charlottesville, Virginia, 2009.

\bibitem{Mitchell1988}
J.~C. Mitchell and G.~D. Plotkin.
\newblock Abstract types have existential type.
\newblock {\em ACM Trans. Program. Lang. Sys.}, 10(3):470--502, 1988.

\bibitem{Moitra1991}
D.~Moitra.
\newblock Finding a minimal cover for binary images: An optimal parallel
  algorithm.
\newblock {\em Algorithmica}, 6(1-6):624--657, June 1991.

\bibitem{Ofman1963}
Y.~Ofman.
\newblock On the algorithmic complexity of discrete functions.
\newblock {\em Sov. Phys. Dokl.}, 7(7):589--591, 1963.

\bibitem{Reinders2007}
J.~Reinders.
\newblock {\em Intel threading building blocks: outfitting {C}++ for multi-core
  processor parallelism}, volume 2007.
\newblock O'Reilly, Sebastopol, CA, 2007.

\bibitem{Sanders2006}
P.~Sanders and J.~L. Tr\"{a}ff.
\newblock Parallel prefix (scan) algorithms for {MPI}.
\newblock In B.~Mohr, J.~L. Tr\"{a}ff, J.~Worringen, and J.~Dongarra, editors,
  {\em Recent Advances in Parallel Virtual Machine and Message Passing
  Interface}, Lecture Notes in Computer Science, chapter~15, pages 49--57.
  Springer, Berlin, Heidelberg, 2006.

\bibitem{Scott1976}
D.~Scott.
\newblock Data types as lattices.
\newblock {\em SIAM J. Comput}, 5(3):522--87, 1976.

\bibitem{Sengupta2007}
S.~Sengupta, M.~Harris, Y.~Zhang, and J.~D. Owens.
\newblock Scan primitives for {GPU} computing.
\newblock In {\em GH '07 Proceedings of the 22nd ACM SIGGRAPH / EUROGRAPHICS
  symposium on Graphics hardware}, pages 97--106, Aire-la-Ville, Switzerland,
  2007. Eurographics Association.

\bibitem{Shah2013}
V.~B. Shah, A.~Edelman, S.~Karpinski, and J.~Bezanson.
\newblock Novel algebras for advanced analytics in julia.
\newblock In {\em 2013 IEEE High Performance Extreme Computing Conference
  (HPEC)}, pages 1--4, Waltham, MA, 2013. IEEE.

\bibitem{Sklansky1960}
J.~Sklansky.
\newblock Conditional-sum addition logic.
\newblock {\em IEEE Trans. Electronic Comput.}, EC-9(2):226--231, 1960.

\bibitem{Smith1987}
D.~R. Smith.
\newblock Applications of a strategy for designing divide-and-conquer
  algorithms.
\newblock {\em Sci. Comput. Program.}, 8(3):213--229, 1987.

\bibitem{Snir1995}
M.~Snir, S.~Otto, S.~Huss-Lederman, D.~Walker, and J.~Dongarra.
\newblock {\em {MPI}: The Complete Reference}.
\newblock MIT Press, Cambridge, MA, 1995.

\bibitem{Strachey2000}
C.~Strachey.
\newblock Fundamental concepts in programming languages.
\newblock {\em Higher-Order Symbol. Comput.}, 13(1-2):11--49, 2000.

\bibitem{Tait1965}
W.~W. Tait.
\newblock Infinitely long terms of transfinite type.
\newblock In J.~N. Crossley and M.~A.~E. Dummett, editors, {\em Formal Systems
  and Recursive Functions}, volume~40 of {\em Studies in Logic and the
  Foundations of Mathematics}, chapter~10, pages 176--185. North-Holland
  Publishing, Amsterdam, 1965.

\bibitem{Viswanathan1996}
G.~Viswanathan and J.~R. Larus.
\newblock {User-defined Reductions for Efficient Communication in Data-Parallel
  Languages}.
\newblock Technical report, University of Wisconsin, Madison, 1996.

\bibitem{Wang1996}
H.~Wang, A.~Nicolau, and K.-Y.~S. Siu.
\newblock The strict time lower bound and optimal schedules for parallel prefix
  with resource constraints.
\newblock {\em IEEE Trans. Comput.}, 45(11):1257--1271, 1996.

\bibitem{Wilt2013}
N.~Wilt.
\newblock Scan.
\newblock In {\em The CUDA Handbook: A Comprehensive Guide to GPU Programming},
  chapter~13. Pearson Education, 2013.

\end{thebibliography}

\appendix

\section{The \code{render} function}
\label{sec:render}

This appendix shows the \code{render} function used to generate the figures in
Figure~\ref{fig:gates}. We use here the Compose package for declarative vector
graphics in Julia~\cite{Compose.jl}.

\code{render} is defined with two methods. The first describes how to render
each elementary operation is represented as a logic \code{gate} with inputs
\code{ins} and outputs \code{outs}. The \code{render(\allowbreak{}G::gate, ...)} method
draws the inputs as small circles at coordinates \code{ipoints} and links them
to the outputs, which are drawn as large circles at coordinates \code{opoints}.
The second method for \code{render} describes how to render the instrumentation
data in an \code{AccessArray}: a first pass through the data computes the depth
of the tree to draw, and the second pass actually places gates appropriately
for each operation, and finally vertical guidelines for every processor are
added.

The code as written uses a heuristic taking advantage of the sequential
left-to-right access order of the serialized prefix sum kernels: if an
operation accesses an index lower than the most recently accessed index, then
the current operation defines a new layer of the computation tree. This
simplifying assumption does not fundamentally change the main idea of rendering
instrumentation data being acquired at the level of individual variables, and
more sophisticated reasoning about tree depths can be used as necessary.

\begin{verbatim}
using Compose

type gate
    ins :: Vector
    outs:: Vector
end

function render(G::gate, x, y, y0; ri=0.1, ro=0.25)
    ipoints = [(i, y0+ri) for i in G.ins]
    opoints = [(i, y0+0.5) for i in G.outs]
    igates  = [circle(i..., ri) for i in ipoints]
    ogates  = [circle(i..., ro) for i in opoints]
    lines = [line([i, j]) for i in ipoints,
                              j in opoints]

    compose(context(units=UnitBox(0.5, 0, x, y+1)),
        compose(context(), stroke("black"),
	    fill("white"), igates..., ogates...),
        compose(context(), linewidth(0.3mm),
	    stroke("black"), lines...))
end

function render(A::AccessArray)
    #Scan to find maximum depth
    olast = depth = 0
    for y in A.history
        (any(y[1] .<= olast)) && (depth += 1)
        olast = maximum(y[2])
    end
    maxdepth = depth
    
    #Map each operation onto a gate
    olast = depth = 0
    C = Any[]
    for y in A.history
        (any(y[1] .<= olast)) && (depth += 1)
        push!(C, render(gate(y...), A.length,
	    maxdepth, depth))
        olast = maximum(y[2])
    end
    
    #Compose everything together with processor
    #guidelines
    push!(C, compose(context(
      units=UnitBox(0.5, 0, A.length, 1)),
      [line([(i,0), (i,1)]) for i=1:A.length]...,
      linewidth(0.1mm), stroke("grey")))
    compose(context(), C...)
end
\end{verbatim}

\end{document}